\documentclass[11pt]{article}
\usepackage{graphicx}
\usepackage{amssymb}
\usepackage{amsmath}
\usepackage{dcolumn}
\usepackage{epstopdf}
\usepackage{bm}
\usepackage{cite}
\title{Low Energy Effective Hamiltonian for a Periodic Array of Cosmological Branes : The Smectic Universe}
\author{D. R. Daniels\footnote{email ddd@aber.ac.uk}}
\begin{document}
\maketitle
\parindent=0pt\vskip-0.5cm
\centerline{Institute of Mathematical and Physical Sciences, University of Wales Aberystwyth, Aberystwyth SY23 3BZ, UK.}

\begin{abstract}
We investigate a simple model of a stack of four dimensional cosmological branes in a five dimensional flat bulk via the use of a derived low-energy effective Hamiltonian for a `radion'-like field associated with brane displacements and deformations. We also extend by analogy the theory of $3D$ smectic liquid crystals and multi-lamellar amphiphilic membranes, as are typically found in soft condensed matter systems, to the $5D$ case of many parallel brane universes. The underlying rotational invariance of such a system implies the absence of an explicit intra-brane cosmological constant term in the low-energy effective Hamiltonian. The strength of a quadratic inter-brane potential introduced for stabilisation is calculated self-consistently, and leads naturally to a novel exponential suppression mechanism for the $4D$ intra-brane cosmological constant. We also pursue, again by analogy to the $3D$ case, the link between this work and the Abelian Higgs transition, or superconductivity, in $5D$. The relevance of the work presented here in explicating the cosmological constant problem is also highlighted, outlined and discussed.


\end{abstract}

\section*{Introduction}

In this work, we consider a simplified model of a periodic array of cosmological branes, which depends on how four dimensional branes are embedded in a flat five dimensional Euclidean bulk. We derive the low-energy effective Hamiltonian for a `radion-like' scalar field $\phi$ which governs the fluctuations in the inter-brane distance between successive branes around some equilibrium brane separation distance $d$. We find that to quadratic order in $\phi$ gravitational effects described by the more familiar Einstein-Hilbert  term do not contribute to the effective Hamiltonian, but that rather the effects of the induced curved brane geometry are determined by the extrinsic curvature tensor. While of course being inspired by them, our simple model is therefore somewhat different to other current brane models \cite{{randall},{arkani1},{dvali},{gregory},{brax1},{dick},{durrer}}. Moreover, although the crude model to be presented below seems perhaps to be a little ad hoc and possibly overly phenomenological in appearance in some respects, it is nevertheless physically well-motivated, and is the simplest kind of model which leads to the interesting  and novel suppression mechanism for the intra-brane cosmological constant as found in this work.

Furthermore, there has been a great deal of cross-fertilisation over the years between what may be traditionally termed `hard' condensed matter physics and particle physics, as given for example by the theory of phase transitions, spontaneous symmetry breaking, and the Higgs phenomenon \cite{{zinn},{lubensky1}}. There has been somewhat less overlap up to now between what is commonly termed `soft' condensed matter physics \cite{{lubensky1}} and high-energy physics.

However, it has recently been proposed that the universe we inhabit is a four dimensional brane embedded in a higher dimensional space \cite{{randall},{arkani1},{dvali},{gregory},{brax1},{dick},{durrer}}. With already a vast literature existing on this topic, such brane models (initially inspired by string theories \cite{{polchinski}}) are thought to shed some light on the cosmological constant problem and dark energy \cite{{weinberg},{brax2},{padma},{peebles},{turner},{riess},{pelmut}}. Variants of these brane models call for the typical structure of two or more such branes, one on top of the other, to be suitably arranged within the higher dimensional space \cite{{arkani2},{hatanaka},{nam},{li},{corley},{deffayet}}.

Furthermore, such an arrangement of membranes has its corresponding lower-dimensional analog in the $3D$ `smectic' (from the greek `$\sigma \mu \epsilon \gamma \mu \alpha$' meaning `soap') phases of amphiphilic molecules or liquid crystals that often occur in soft condensed matter physics (for comprehensive introductions and reviews, see \cite{{lubensky1},{peierls},{landau1},{safran},{nelson},{degennes1},{kamien1},{kamien2},{porte},{dejeu},{kleinert1},{kleinert2}}). The basic brane setup is as shown in Figure~1. Furthermore, the analogy between a soft condensed matter membrane and a cosmological brane has also recently been pursued (albeit in a rather different context from the work presented here) in \cite{{sinha}}. Moreover, one of the main motivations behind the work presented here is to introduce the physics of smectic systems to the wider high-energy community, within the context of cosmological branes. As with all analogies, that between soft matter membranes and cosmological branes has its strengths and weaknesses, with perhaps its greatest strength lying in its heuristic properties.

The work presented here can therefore be simply viewed on one level as the extension of theoretical work familiar from $3D$ smectics to $5D$ branes. Moreover, the general brane picture outlined in this work qualitatively resembles previous work carried out on cosmological branes, such as in \cite{{turok1},{turok2}}, for example. Furthermore, one of the main hopes of this work is that physics borrowed from $3D$ smectics might prove to be a rather fruitful and useful addition to the already vast literature on cosmological and high-energy $5D$ branes.

The other main motivation behind this work lies in the observation that for many $3D$ smectic systems the membranes are more or less tensionless (i.e. at zero surface tension). The $5D$ analog of $3D$ surface tension is the intra-brane cosmological constant. Therefore, the work presented here should shed some light on the cosmological constant problem and dark energy \cite{{weinberg},{brax2},{padma},{peebles},{turner},{riess},{pelmut}}. In our view, one of the difficulties of the cosmological constant problem is that it is a low-energy effect - and as such it is hard to see how to change theories of gravity too much in the infra-red limit without fouling up the previous outstanding successes of General Relativity. One way out of such a perceived dilemma may lie in consideration of similar approaches as that undertaken in the work presented below.

Note that, unless otherwise stated, we will work throughout for the theory presented here in Euclidean space, while assuming that one can always simply Wick-rotate back to Lorentzian spacetime if required at a later date, or whenever necessary.

\vspace{.5in}
\centerline {
\includegraphics[width=4in]{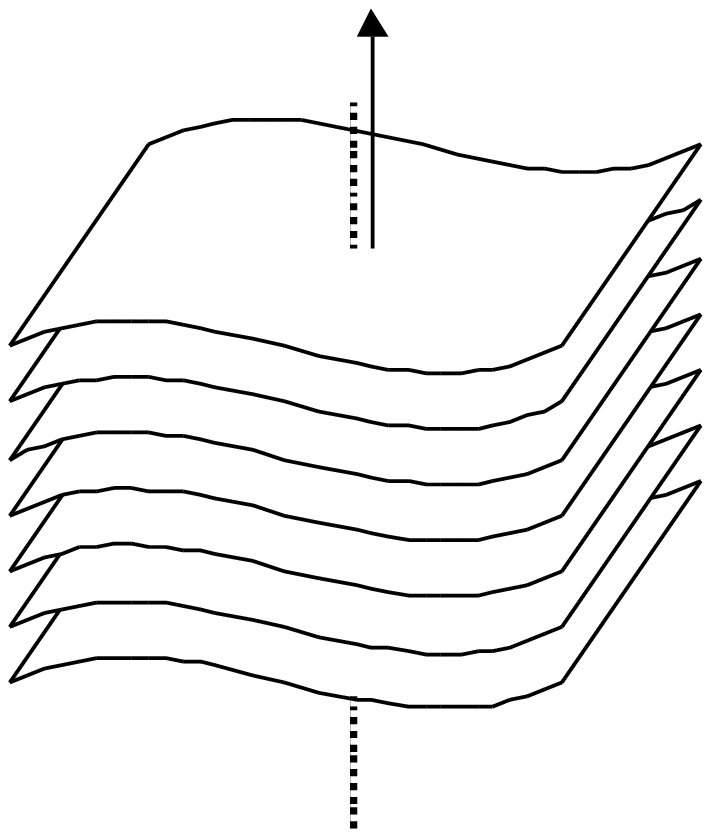}
}
{\bf Fig~1} {\small Sketch of the smectic universe model. Each brane corresponds to a four dimensional universe embedded in a larger, flat, five dimensional universe. The inter-brane spacing is governed by a fluctuating scalar field $\phi$. The equilibrium (ground-state) configuration consists of flat, equidistant layers. Also shown (as a solid directed arrow) is the normal vector of a typical brane.}
\vspace{.5in}

\section*{Embedding Geometry of Smectic Branes}

In this section we investigate the extrinsic geometry of embedding a stack of curved, four-dimensional branes in a flat five-dimensional bulk universe \cite{{akama},{dahia},{shiro},{pavsic},{gogber},{maia1},{maia2},{maia3},{maia4}}. We embed the position of the $n$th brane, $r_n^A (x)$, as follows (where the index $A$ runs from $1 \ldots 5$, the index $\mu$ runs from $1 \ldots 4$, and the index $n$ runs from $-\infty \ldots +\infty$) :

\begin{equation}
r_n^A (x) = x^{\mu} \delta_{\mu}^A + X_n (x) \delta_5^A
\label{multi1}
\end{equation}

How to parameterise the embedding is somewhat arbitrary. The choice given in Eq~(\ref{multi1}) corresponds to choosing the higher dimensional analog of the `Monge gauge', as used often when dealing with two dimensional membranes \cite{{safran},{nelson},{kamien2}}. It thus follows that the induced metric on brane $n$ becomes :

\begin{equation}
g_{\mu \nu}^n = \delta_{\mu \nu} + \partial_{\mu} X_n \partial_{\nu} X_n
\label{multi2}
\end{equation}

while the unit normal vector to the $n$th brane becomes :

\begin{equation}
N_n^A = \frac{\delta_5^A - \partial_{\mu} X_n}{\sqrt{g_n}}
\label{multi3}
\end{equation}

The extrinsic curvature of layer $n$ becomes :

\begin{equation}
K_{\mu \nu}^n = \delta_{A B} N_n^A \partial_{\mu} \partial_{\nu} r_n^B
\label{multi4}
\end{equation}

Additionally, the extrinsic and intrinsic curvatures are related via the well known Gauss-Codazzi relation \cite{{akama},{dahia},{shiro},{pavsic},{gogber},{maia1},{maia2},{maia3},{maia4}} :

\begin{equation}
{\cal{R}} = R + K^2 - K_{\mu \nu} K^{\mu \nu}
\label{multi4a}
\end{equation}

where ${\cal{R}}$ and $R$ are the five dimensional and four dimensional Ricci curvature scalars respectively, and $K = K_{\mu}^{\mu}$ is the trace of the extrinsic curvature tensor. We consider in this work a flat, five dimensional, Euclidean bulk, and so ${\cal{R}} = 0$. Using the Gauss-Codazzi relation of Eq~(\ref{multi4a}), we can therefore speak in terms of either the intrinsic intra-brane curvature $R$ or the extrinsic brane curvature $K_{\mu \nu}$ interchangeably.

The covariant distance between consecutive branes along the layer normal (see Figure~2) is given by $\Delta_n$ :

\begin{eqnarray}
\Delta_n & = &  \delta_{A B} ( r_{n+1}^A - r_n^A ) N_n^B \nonumber \\
& = & \frac{X_{n+1} - X_n}{\sqrt{g_n}}
\label{multi4b}
\end{eqnarray}

In what follows we will find it useful to assume that the ground-state multi-brane configuration is given by almost flat and equidistantly separated membranes, such that $X_n = n d + \phi_n (x)$, where $d$ is the equilibrium inter-brane separation distance. We can then perform a perturbative expansion in $\phi_n$ for all quantities of interest. In this way, we find that : 

\begin{eqnarray}
g_{\mu \nu}^n & \approx & \delta_{\mu \nu} + \partial_{\mu} \phi_n \partial_{\nu} \phi_n \nonumber \\
N_n^A & \approx & \big( 1- \frac{1}{2} ( \partial_{\mu} \phi )^2 \big) \delta_5^A - \partial_{\mu} \phi_n \nonumber \\
K_{\mu \nu}^n & \approx & \partial_{\mu} \partial_{\nu} \phi_n \nonumber \\
\Delta_n & \approx & \big( 1- \frac{1}{2} ( \partial_{\mu} \phi )^2 \big) d + ( \phi_{n+1} - \phi_n )
\label{multi4c}
\end{eqnarray}

to quadratic order in $\phi_n$.

\vspace{.5in}
\centerline {
\includegraphics[width=4in]{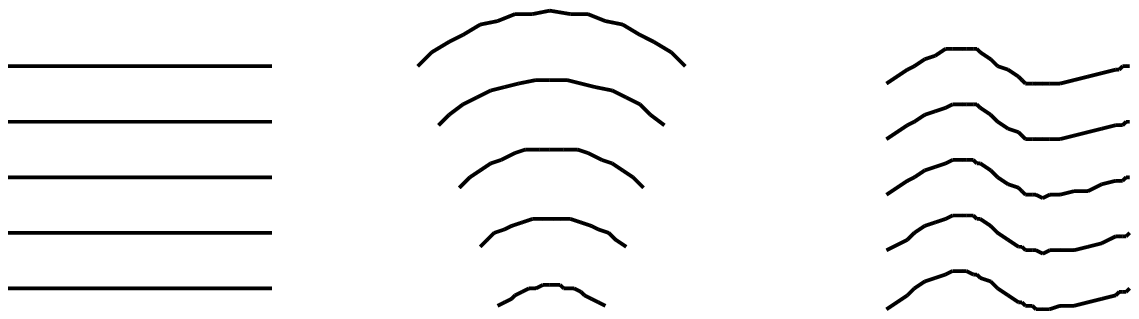}
}
{\bf Fig~2} {\small On the left we have depicted the smectic universe ground-state as equally spaced, flat branes. In the middle, and on the right we have depicted deformations of the ground-state smectic universe which maintain the equal spacing distance $d$ of each brane (along the normal vector $N_n^A$ of each brane). However, unlike the figure on the far left, the smectic universes shown in the middle and on the far right are non-flat and thus possess mean curvature $K_n$. There is a concomitant energy cost associated with such a bending curvature of the branes.}
\vspace{.5in}

\section*{Rotational Invariance}

The layering of the smectic universe spontaneously breaks the translational as well as rotational symmetry of the underlying bulk spacetime. As such, any variation of the field $\phi$ that corresponds to a rigid rotation of the brane layers must leave the Hamiltonian invariant (see Figure~3). Therefore distortions of the smectic universe that leave the inter-brane layer separation distance unchanged (and therefore the Hamiltonian) are dominant (For details of the analogous and corresponding treatment of $3D$ smectics see \cite{{lubensky1},{degennes1},{golubovic},{holyst},{grinstein},{pelcovits},{landau2},{fournier}}).

An infinitesimal global rotation of the brane layers would lead to a transformation on $\phi$ of :

\begin{equation}
\phi \rightarrow \phi + \Omega_{\mu} x^{\mu}
\label{rot1}
\end{equation}

such that the layer normal $N^A$ according to Eq~(\ref{multi3}) gets transformed in accordance with :

\begin{equation}
\partial_{\mu} \phi \rightarrow \partial_{\mu} \phi + \Omega_{\mu}
\label{rot2}
\end{equation}

which means that $\partial_{\mu} \phi$ is simply not rotationally invariant. We can thus see that due to the underlying rotational invariance there can be no term in the low-energy effective Hamiltonian corresponding to $\big( \partial_{\mu} \phi \big)^2$. Furthermore, it is precisely this rotational symmetry argument that is also the reason why no explicit intra-brane cosmological constant term can appear in the low-energy effective Hamiltonian. Such an explicit cosmological constant term would necessarily lead to a contribution in our approximation of : $\sqrt{g} \approx 1 + \frac{1}{2} \big( \partial_{\mu} \phi \big)^2$ to the low-energy effective Hamiltonian, and would thus be disallowed due to rotational symmetry grounds.

The smectic universe can thus be seen as a phase transition to a layered $4D$ brane configuration from a bulk $5D$ space, with an associated symmetry breaking of rotational and translational symmetry.

\vspace{.5in}
\centerline {
\includegraphics[width=4in]{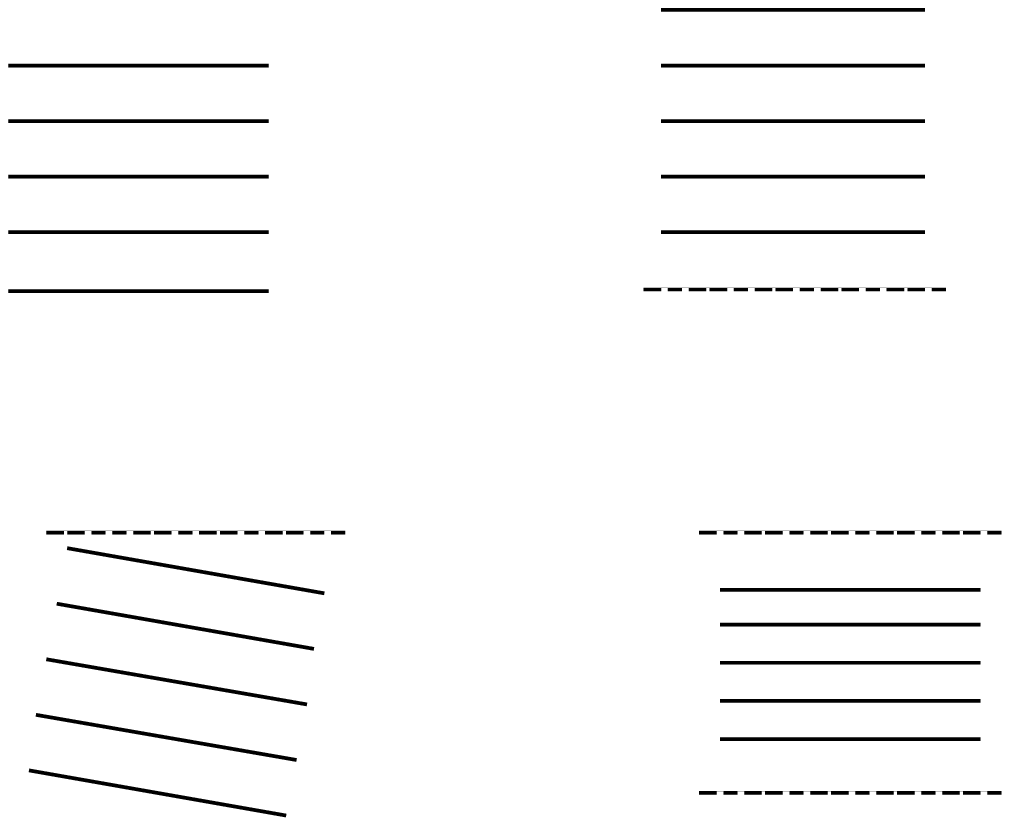}
}
{\bf Fig~3} {\small The top left diagram depicts the ground-state smectic universe with equilibrium inter-brane spacing. The top right diagram shows a simply translated smectic universe, with the same energy as the ground-state. The bottom left diagram depicts a simply rotated smectic, with the same energy as the ground-state. The bottom right diagram shows a smectic universe with compressed inter-brane spacing, and therefore an energy greater than the ground-state configuration. (The dashed lines represent the reference brane ground-state of equally spaced, flat layers.)}
\vspace{.5in}

\section*{Heuristic Derivation of the Low-Energy Effective Hamiltonian}

We now outline a much simplified derivation of the low-energy effective Hamiltonian for a periodic array of branes, for heuristic purposes, and in order to motivate the following sections.

We parameterise our theory in the full five dimensional bulk via the `radion-like' field $\phi$ which describes the deviation of branes from their ideal, equally spaced and flat configurations. Given the underlying translational invariance of our theory, there can be no mass term (proportional to $\phi^2$) in our Hamiltonian. Next to consider are terms with two derivatives : $\big( \partial_5 \phi \big)^2$, roughly perpendicular to the branes along the $x^5$ direction, and $\big( \partial_{\mu} \phi \big)^2$ along the branes. Due to rotational invariance, we have seen above that there can be no term in the low-energy effective Hamiltonian proportional to $\big( \partial_{\mu} \phi \big)^2$. Therefore in order to describe the effects of $\phi$ along the branes we need to go to next order in derivatives : $\big( \partial_{\mu}^2 \phi \big)^2$. Such a term, as we have also seen from above, arises due to membrane extrinsic curvature. So, to lowest order in $\phi$ and its derivatives, the only terms which contribute to the low-energy effective Hamiltonian are those given by : $\big( \partial_5 \phi \big)^2$ and $\big( \partial_{\mu}^2 \phi \big)^2$.

\section*{Analogous `Spin-Wave' Hamiltonian}

For further heuristic purposes, we can consider the following five dimensional Hamiltonian inspired by the lower dimensional physics of layering phenomena in `spin-waves' \cite{{grinstein},{pelcovits}} at some density $\rho$ :

\begin{equation}
{\cal H}_{\rho}  = \frac{1}{2} \int d^5 x \Big( q_0^4 | \rho |^2 - 2 q_0^2 | \partial_A  \rho |^2  + | \partial_A^2  \rho |^2 \Big)
\label{spin1}
\end{equation}

where we can think of $q_0 = 2 \pi / d$ as representing the inter-brane spacing. The Hamiltonian must be translationally invariant, and so no mass term proportional to $\rho^2$ can appear in Eq~(\ref{spin1}). Since the coefficient of $| \partial_A  \rho |^2$ is negative, the underlying rotational symmetry is broken (as outlined in \cite{{lauscher}}), accompanied by the formation of smectic-type layers. In this broken phase, we can write for the `spin-wave' field, $\rho = \rho_0 \exp \big( i q_0 (x^5 + \phi ) \big)$, which represents a slow, layered modulation of $\rho$ along the $x^5$ direction. Substituting this ansatz into Eq~(\ref{spin1}), we get as the ground-state configuration, to lowest order in $\phi$ and its derivatives :

\begin{equation}
{\cal H}_{\rho}  = 2 q_0^4 | \rho_0 |^2 \int d^5 x  \Big( \big( \partial_5 \phi \big)^2 + \frac{1}{4 q_0^2} \big( \partial_{\mu}^2 \phi \big)^2 \Big)
\label{spin2}
\end{equation}

One may be somewhat surprised by the anisotropy present in Eq~(\ref{spin2}). However, the formation of smectic layers spontaneously breaks the rotational (and translational) symmetry of the system. As such, any variation of the field $\phi$ that corresponds to a rigid rotation of the brane layers must leave the Hamiltonian (Eq~(\ref{spin2})) invariant (see also Figure~3).

\section*{Smectic Branes Hamiltonian}

From the considerations and results of above, we can now therefore, in complete analogy with the lower-dimensional $3D$ liquid crystal case, write the total $5D$ smectic Hamiltonian summed over all branes as :

\begin{equation}
{\cal{H}} = \frac{1}{2} \sum_n \int d^4 x \sqrt{g_n} \Big( \kappa K_n^2  + B \big( \Delta_n - d \big)^2 \Big)
\label{multi5}
\end{equation}

$\kappa$ and $B$ are constants introduced in order to measure the relative strength of the contribution to the Hamiltonian from the intra-layer bending and the inter-layer compressional interaction respectively.

In Eq~(\ref{multi5}) the energy cost associated with brane layer bending means that the mean extrinsic curvature, $K_n$, is penalised, and the branes prefer to remain as flat as possible. The reason for the use of such a term (proportional to $K_n^2$) in the above Hamiltonian, as opposed to say that of a more familiar Einstein-Hilbert term will be outlined below. Furthermore, such a term (proportional to $K^2$) also analogously arises in some string theory model Hamiltonians \cite{{polyakov},{david},{kleinert3}}.

The quadratic inter-brane potential in Eq~(\ref{multi5}) represents the simplest and most convenient potential one could write down (with no loss of generality), for branes around some equilibrium brane separation $d$, and characterised by the quadratic strength $B$. Furthermore, we will calculate the potential strength $B$ self-consistently later, using a theoretical approach \cite{{helfrich},{seifert}} that has found great success in soft condensed matter systems.

We can see that no explicit (or isolated) cosmological constant terms appear in Eq~(\ref{multi5}), but only as in the combination shown, including $\Delta_n$. As we have seen, this is ultimately due to the underlying rotational invariance of a smectic arrangement of branes (for details of the analogous case in $3D$ see \cite{{lubensky1},{degennes1},{golubovic},{holyst},{grinstein},{pelcovits},{landau2},{fournier}}).

Additionally, by performing a small $\Delta_n$ expansion in Eq~(\ref{multi5}), we can easily read-off the relevant cosmological constants, such that we must identify (term by term) : $\frac{2 \Lambda_4}{16 \pi G} = \frac{B d^2}{2}$ and $\frac{2 \Lambda_5}{16 \pi G} = - B d$, where $\Lambda_4$ and $\Lambda_5$ are the intra-brane and inter-brane cosmological constants respectively. We shall have further recourse to these results below.

Another, more heuristic way of seeing how to obtain the inter-brane potential of Eq~(\ref{multi5}) is to write it as : $B \Delta_n^2 + \frac{2 \Lambda_5}{16 \pi G} \Delta_n + \frac{2 \Lambda_4}{16 \pi G}$, where again $\Lambda_4$ and $\Lambda_5$ are the intra-brane and inter-brane cosmological constants respectively. Rearranging we get : $\frac{B}{2} \big( \Delta_n + \frac{2 \Lambda_5}{16 \pi G B} \big)^2 + \frac{2 \Lambda_4}{16 \pi G} - \frac{1}{2 B} \big( \frac{2 \Lambda_5}{16 \pi G} \big)^2$, such that we identify $\frac{2 \Lambda_5}{16 \pi G B} = - d$, and as we shall see later, $\frac{2 \Lambda_4}{16 \pi G} - \frac{1}{2 B} \big( \frac{2 \Lambda_5}{16 \pi G} \big)^2$ must vanish due to the underlying smectic rotational invariance.

We now proceed from Eq~(\ref{multi5}) by assuming that the ground-state multi-brane configuration is given by almost flat and equidistantly separated membranes, such that $X_n = n d + \phi_n (x)$. Expanding perturbatively in $\phi_n (x)$, and using the results of Eq~(\ref{multi4c}), we can simplify the Hamiltonian (\ref{multi5}) further so that it becomes :

\begin{equation}
{\cal{H}} = \frac{1}{2} \sum_n \int d^4 x \Big( \kappa \big( \partial_{\mu}^2 \phi_n \big)^2  + B \big( \phi_{n+1} - \phi_n \big)^2 \Big)
\label{multi6}
\end{equation}

Moreover, under this approximation, we can show that any contribution from a more familiar Einstein-Hilbert term (proportional to $R$) in Eq~(\ref{multi6}) vanishes. Using the Gauss-Codazzi relation of Eq~(\ref{multi4a}), (with ${\cal{R}}=0$), we can always rewrite $R$ in terms of $K_{\mu \nu}$. Furthermore, using the results of Eq~(\ref{multi4c}), we can see that to lowest order in $\phi_n$, it must follow that any contribution that goes like $\int d^4x R$ in the effective Hamiltonian must vanish (up to surface terms). This explains the absence of any canonical Einstein-Hilbert term in our original Hamiltonian, Eq~(\ref{multi5}) - such a term simply does not contribute to ${\cal{O}} ( \phi_n^2 )$.

Proceeding further, as long as we probe distance scales along the $5$ direction greater than $d$, we can safely pass to the continuum limit and write our Hamiltonian, with $x^5 = n \, d$, as :

\begin{equation}
{\cal{H}} = \frac{1}{2 d} \int d^5 x \Big( \kappa \big( \partial_{\mu}^2 \phi \big)^2  + B d^2 \big( \partial_5 \phi \big)^2 \Big)
\label{multi7}
\end{equation}

which corresponds to the higher-dimensional $5D$ analog of the $3D$ smectic phase familiar from amphiphilic membranes and liquid crystals of soft condensed matter physics \cite{{lubensky1},{peierls},{landau1},{safran},{nelson},{degennes1},{kamien1},{kamien2},{porte},{dejeu},{kleinert1},{kleinert2}}.

Thus Eq~(\ref{multi7}) represents the low-energy, effective Hamiltonian of the `radion'-like field $\phi$ characterising a stack of branes arranged in a $5D$ bulk universe. The inter-brane interaction is quadratic and attractive, whose strength is governed by a single parameter $B$ and leads to an equilibrium separation distance $d$ between consecutive branes in the stack.

Note again the anisotropy present in Eq~(\ref{multi7}). The layering of the smectic universe spontaneously breaks the translational as well as rotational symmetry of the underlying bulk spacetime. Therefore, any variation of the field $\phi$ that corresponds to a rigid rotation of the brane layers must leave the Hamiltonian invariant (as shown in Figure 3).

\section*{Self-Consistent Calculation of Potential Strength ($B$)}

The quadratic inter-brane potential in Eq~(\ref{multi5}) represents the simplest potential one could write down (with no loss of generality), expanded around some equilibrium brane separation $d$, and characterised by the quadratic strength $B$. We now proceed to calculate $B$, via the five-dimensional treatment of that analogously given for $3D$ membranes in \cite{{safran},{helfrich},{seifert}}.

The self-consistent condition that must be met by the strength of the inter-brane potential, as governed by $B$, is given such that typically (and approximately) on average : 

\begin{equation}
< \big( \phi_{n+1} - \phi_n \big)^2 > \, = \,  d^2
\label{B1}
\end{equation}

where the averaging $< \ldots >$ is to be carried out using Eq~(\ref{multi7}). This translates in the continuum limit (provided we use a cut-off of $\pi / d$ for momenta along $x_5$) into the condition :  $< \big( \partial_5 \phi \big)^2 > \, = \, 1 $. This condition is required in order for the quadratic expansion in $\Delta_n$ around equally spaced branes for the inter-brane potential, as used in Eq~(\ref{multi5}), to be self-consistent. Moreover, if the condition of Eq~(\ref{B1}) were not met then the branes would become strongly-interpenetrating, the notion of individual, equally spaced, and well-separated membranes would break down, and the consistency of our calculational scheme would become violated.

Therefore, using Eq~(\ref{multi7}) to find the propagator for $\phi$, the equation we need to solve for $B$ is given by :

\begin{equation}
\int_0^{\pi / d} \frac{dp}{\pi} \int \frac{d^4 q}{( 2 \pi )^4} \frac{p^2 d}{\kappa q^4 + B d^2 p^2} \, = \, 1
\label{B2}
\end{equation}

Introducing an additional intra-brane small distance cut-off $\epsilon$, such that $\int d^4 q \rightarrow 2 \pi^2 \int_0^{\pi  / \epsilon} q^3 dq$, and solving for $B$, as $\epsilon \rightarrow 0$, we get :

\begin{equation}
B=\frac{\kappa \pi^2}{\epsilon^4} \exp \big( - 96 \kappa d^2 + 2/3 \big)
\label{B3}
\end{equation}

The property of interest to immediately note from Eq~(\ref{B3}) is the exponential suppression of $B$ as $96 \kappa d^2 \gg 1$. This strong suppression effect is unique and peculiar to the smectic brane arrangement considered in this work. It is ultimately due to the presence of four derivatives in the original Hamiltonian (\ref{multi7}), describing propagation of $\phi$ on four-dimensional branes. This effect can also be seen heuristically and on dimensional grounds as being due to the appearance of $\sim \int d^4 q / q^4$ in the $\phi$ field propagator.

\section*{Intra-Brane Cosmological Constant (`Surface Tension')}

Using the results of above, and via inspection of Eq~(\ref{multi5}), we are naturally led to identify $\kappa = \frac{1}{8 \pi G} = \frac{1}{8 \pi l_p^2}$, where $l_p$ is the Planck length (in units where $c = \hbar = 1$). Additionally, it is also natural to identify the intra-brane small-distance cut-off with the Planck length, such that $\epsilon = l_p$. Using the result of Eq~(\ref{B3}) and via inspection of Eq~(\ref{multi5}) as outlined previously above, we can read off the value of the intra-brane cosmological constant, $\Lambda_4$, as follows :

\begin{equation}
\Lambda_4 = 4 \pi l_p^2 B d^2 = \frac{\pi^2 d^2}{2 l_p^4} \exp \big( - \frac{12 d^2}{\pi l_p^2} + \frac{2}{3} \big)
\label{st1}
\end{equation}

Note that it then follows that the intra-brane cosmological constant $\Lambda_4$ becomes exponentially suppressed as $d$ increases above just a few Planck lengths. In order to see just how well this exponential suppression mechanism works, note that experimentally \cite{{padma},{riess},{pelmut}} it is observed that $\Lambda_4 l_p^2 \lesssim 10^{-123}$, which implies that from Eq~(\ref{st1}) we must have $d \gtrsim 9 \, l_p$. Thus, the mechanism of symmetry breaking required for the smectic arrangement of consecutive branes as outlined in this work naturally leads to a vanishingly small effective intra-brane cosmological constant.

\section*{Bulk Cosmological Constant (`Inter-Brane Pressure')}

Similarly and again, using the result of Eq~(\ref{B3}) and via inspection of Eq~(\ref{multi5}) as outlined previously, we can read off the value of the bulk cosmological constant, $\Lambda_5$, as follows :

\begin{equation}
\Lambda_5 = - 8 \pi l_p^2 B d = - \frac{\pi^2 d}{l_p^4} \exp \big( - \frac{12 d^2}{\pi l_p^2} + \frac{2}{3} \big)
\label{p1}
\end{equation}

Note that the inter-brane cosmological constant $\Lambda_5$ also becomes exponentially suppressed as $d$ increases above just a few Planck lengths. Furthermore, from Eq~(\ref{st1}) and Eq~(\ref{p1}), we can see as shown previously above that the relation $\Lambda_4 = - \Lambda_5 \, d / 2$ also holds between the intra-brane and bulk cosmological constants.

\section*{Higgs (`Superconductivity') Analog}

It has long been known that there exists a powerful analogy between the $3D$ phase transition physics associated with smectic liquid crystals and superconductivity \cite{{lubensky1},{degennes1},{degennes2},{halperin},{lubensky2},{lubensky3},{lubensky4}}. The purpose of this section is to pursue this analogy further for the case of $5D$ as considered in this work.

We begin by giving the canonical Hamiltonian \cite{{zinn}} for the Abelian Higgs mechanism in a flat, bulk, five-dimensional Euclidean space :

\begin{equation}
{\cal H}_{TOT} = {\cal H}_{\psi} + {\cal H}_{\delta N} + {\cal H}_{GF}
\label{super1}
\end{equation}

where the `matter' Hamiltonian is given by :

\begin{equation}
{\cal H}_{\psi}  = \frac{1}{2} \int d^5 x \Big( | ( \partial_A - i q_0 \, \delta N_A ) \psi |^2  + \frac{m^2}{2} | \psi |^2 + \frac{\lambda}{4} | \psi |^4 \Big)
\label{super2}
\end{equation}

with the `charge' given in terms of the inter-brane spacing by $q_0 = 2 \pi / d$, and where the `vector potential' is given in terms of the change in the brane normal vector by $\delta N^A = N^A - \delta_5^A$. The `gauge' Hamiltonian is given by the usual Maxwell term :

\begin{equation}
{\cal H}_{\delta N}  = \frac{1}{4} \int d^5 x  F_{AB} F^{AB}
\label{super3}
\end{equation}

with the `field strength' given by $F_{AB} = \partial_A \delta N_B - \partial_B \delta N_A$. As always with gauge theories we also require a gauge-fixing term $ {\cal H}_{GF}$. There is typically much freedom in the choice of a gauge-fixing term one uses. Furthermore, often the best choice of gauge is one that highlights the underlying physics. Moreover, due to the underlying gauge invariance of the Hamiltonian, the physics remains the same whatever the gauge choice. Therefore, for the purposes of this work, we conveniently chose ${\cal H}_{GF}$ to be as follows :

\begin{equation}
{\cal H}_{GF}  = \frac{1}{2 d} \int d^5 x \Big( \kappa ( \partial_{\mu} \delta N^A )^2 + B d^2 ( \delta_A^5 \delta N^A )^2 \Big)
\label{super4}
\end{equation}

When $m^2 < 0$ in Eq~(\ref{super2}), spontaneous symmetry breaking occurs, accompanied by the formation of smectic layers. In this broken phase, we can write for the `matter' field, $\psi = \psi_0 \exp ( i q_0 \phi )$, which represents a slow, layered modulation of the scalar field $\psi$ along the $x^5$ direction. The minimum of Eq~(\ref{super1}) now occurs when $| \psi_0 |^2 = - m^2 / \lambda$ and $\delta N_A = \partial_A \phi$. In this way (albeit in a somewhat unconventional gauge) the Abelian Higgs Hamiltonian ${\cal H}_{TOT} $ of Eq~(\ref{super1}) becomes identical to that of the smectic Hamiltonian as given by Eq~(\ref{multi7}), and the analogy is complete.

\section*{Conclusion}

In this work, we have considered a simplified model of a periodic array of cosmological branes, which depends on the extrinsic curvature of embedded, four dimensional branes in a flat five dimensional bulk. We derived the low-energy effective Hamiltonian for a scalar field $\phi$ which governs the fluctuations in the inter-brane distance between successive branes around some equilibrium brane separation distance $d$. To quadratic order in $\phi$ gravitational effects described by the more familiar Einstein-Hilbert  term do not contribute to the effective Hamiltonian. While of course being inspired by them, our crude model is therefore somewhat different to other current brane models \cite{{randall},{arkani1},{dvali},{gregory},{brax1},{dick},{durrer}}. Moreover, despite the model presented in this work appearing perhaps a little ad hoc and possibly overly phenomenological in some respects, it is nevertheless physically well-motivated, and is the simplest kind of model which leads to the interesting and novel suppression mechanism for the intra-brane cosmological constant as outlined above.

We have also extended by analogy the $3D$ theory of smectic liquid crystals and multi-lamellar amphiphilic membranes, to the $5D$ case of many parallel brane universes, in the hope that physics borrowed from $3D$ smectics might shed some light on cosmological $5D$ branes. As with all analogies, there are strengths and weaknesses in using this approach, with perhaps the greatest strength of the analogy pursued here between soft matter branes and cosmological branes lying in its heuristic properties. Additionally, one main motivation behind this work is the observation that many $3D$ smectic systems exist at roughly zero surface tension. The $5D$ analog of $3D$ surface tension is the intra-brane cosmological constant : $\Lambda_4$. Therefore, it is hoped that one of the main benefits of the approach taken in this work is that it might shed some light on the cosmological constant problem and associated dark energy.

We have shown that rotational invariance implies the absence of an explicit cosmological constant term in the Hamiltonian describing the low-energy fluctuations of a brane in a multi-brane arrangement. In this sense one can say that a symmetry principle constrains the cosmological constant to be vanishingly small. Furthermore, this symmetry principle is additionally likely to be a Ward identity also in $5D$ (when issues relating to the correct measure for path integration are taken into account), as is the case in $3D$ \cite{{golubovic},{halperin}}. It is very rare to theoretically find such a constraint on the cosmological constant due to an underlying symmetry principle. The only other well known case is that given by supersymmetry of course, which we do not discuss in this work.

We have also shown in this work a novel possible relaxation mechanism for producing a vanishingly small intra-brane cosmological constant.  Namely, if the distance between branes is at its equilibrium value, then the induced cosmological constant on the brane must vanish. If the inter-brane separation deviates from its equilibrium value, then the intra-brane cosmological constant must adjust itself accordingly. Put differently, any contribution to the intra-brane cosmological constant (from any source on the brane) can easily be absorbed by a concomitant relaxation of the inter-brane separation distance.

Taken in total, the work presented here goes some way to explicating the cosmological constant problem. In this picture, the big bang corresponds to a collision between two branes, as found in models such as the ekpyrotic \cite{{turok1}} or cyclic \cite{{turok2}} universes for example. Thus the work presented here can be thought of as representing the low-energy, effective physics of the corresponding `radion-like' field describing a stack of branes in a $5D$ bulk universe. The great benefit of the work presented here is that it requires no fine-tuning in order to produce a vanishingly small cosmological constant. A small cosmological constant arises perfectly naturally, and self-consistently in the approach outlined above.

Note that the effects described in this work disappear for an isolated, solitary non-interacting brane, and only arise within the context of a stack of many interacting branes. In this sense, the effective properties of any single, individual brane has its origin in a `many- body' effect, due to the many-brane interactions.

Perhaps the best and simplest way to understand the work presented here is in terms of an expansion in $\phi$ and its derivatives in the low-energy effective Hamiltonian as follows. Due to the vanishing of terms like $\phi^2$ and $\big( \partial_{\mu} \phi \big)^2$ in the effective Hamiltonian, the next order terms in Eq~(\ref{multi7}) behave like $\big( \partial_{\mu}^2 \phi \big)^2$ which involve four derivatives on $\phi$. Given that the effects of $\phi$ then propagate on a $4D$ brane, one is guaranteed (to leading order $\sim \int d^4 q / q^4$) logarithmic behaviour of the $\phi$ field propagator on the brane. It is precisely this feature of the model presented here (peculiar to a theory with four derivatives propagating along four dimensions) that ultimately leads in turn to the novel exponential suppression of the intra-brane cosmological constant for inter-brane separations greater than a few Planck lengths : $d \gtrsim 9 \, l_p$.

It is interesting to note that the results, such as Eq~(\ref{st1}) for $\Lambda_4$, obtained in this work are highly non-analytic in the gravitational coupling constant $G$. This aspect of the work presented here (along with the superconductivity analogy outlined above) is strongly reminiscent of other and very important cases of non-analyticity in a coupling constant that can be found at the heart of phenomena such as the formation of Cooper pairs and the allied superconducting instability in BCS theory \cite{{bardeen}}, as well as the phenomenon of asymptotic freedom in non-Abelian gauge theories \cite{{wilczek}}. In this sense, the work presented here shares some of the same interesting low-energy features as other well-known theories with non-trivial infra-red behaviour.

Finally, higher (anisotropic) derivatives in a Hamiltonian can sometimes spell trouble for a theory with issues of unitarity. However, this is not the case in this work, since we start by describing the physics in a perfectly isotropic bulk. It is only when we constrain the physics of our observable universe to a single, isolated brane that the isotropy gets spontaneously broken. There can therefore be no problems with unitarity for the underlying theory in the bulk. The case considered here is thus inherently different from the more familiar scenarios encountered in quantum field theories, and so the usual unitarity objections sometimes raised, simply do not apply to the work presented here.

\section*{Acknowledgements}

Helpful discussion and correspondence with R.C. Ball, J. Samuel, M.D. Maia, and M. Gogberashvili is most gratefully acknowledged.

\end{document}